\documentclass[aps,twocolumn,showpacs,preprintnumbers,amsmath,amssymb]{revtex4}
\usepackage{graphicx}
\usepackage{dcolumn}
\usepackage{color}

\def\ADD#1{{\textcolor{blue}{#1}}}        

\newcommand{\beq}{\begin{equation}}
\newcommand{\eeq}{\end{equation}}
\begin{document}

\title{\ADD{}Energy and enstrophy dissipation in steady state 2-d turbulence}

\author{Alexandros Alexakis}
\affiliation{National Center for Atmospheric Research\\
             P.O. Box 3000, Boulder, CO 80307-3000, U.S.A. }
\author{Charles R. Doering}
\affiliation{Department of Mathematics and Michigan Center for Theoretical Physics\\
University of Michigan, Ann Arbor, MI 48109-1109, U.S.A.
}

\date{\today}

\begin{abstract}
Upper bounds on the bulk energy dissipation rate $\epsilon$ and enstrophy dissipation rate
$\chi$ are derived for the statistical steady state of body forced two
dimensional turbulence in a periodic domain.
For a broad class of externally imposed body forces it is shown that
$\epsilon \le k_{f}    U^3 Re^{-1/2}\left(C_1+C_2 Re^{-1}\right)^{1/2}$ and 
$\chi     \le k_{f}^{3}U^3          \left(C_1+C_2 Re^{-1}\right)$
where $U$ is the root-mean-square velocity, $k_f$ is a wavenumber (inverse length scale)
related with the forcing function, and $Re = U /\nu k_f$.
The positive coefficients $C_1$ and $C_2$  are uniform in the the kinematic viscosity $\nu$,
the amplitude of the driving force, and the system size.
We compare these results with previously obtained bounds for
body forces involving only a single length scale, or for
velocity dependent a constant-energy-flux forces acting at finite wavenumbers.
Implications of our results are discussed.
\end{abstract}
\maketitle

\section{Introduction}

The study of two dimensional turbulence was originally justified as a simplified version of
the more complex three dimensional turbulence, but it has come to be regarded as an interesting
research field in its own right with deep connections to geophysical and astrophysical problems
such as as strongly rotating stratified flows \cite{Tabeling02}. 
A large number of experimental methods have been devised to constrain flows in two dimensions
(e.g. soap films) allowing some theories theories to be tested in the lab \cite{Kellay02}. 
Direct numerical simulations are far easier than the three dimensional case, and this has enabled
researchers to investigate two dimensional turbulence computationally at much higher 
Reynolds numbers \cite{Pouquet75,Boffeta00,Tran04,Chen03,Danilov05,Dmitruk05}.
As a result, there are more simulation data to test theories of two dimensional turbulence.
Nevertheless many fundamental questions remain open;
see \cite{Tabeling02} for a recent review. 

The inviscid conservation of enstrophy as well as energy in two dimensions results in two 
cascading quadratic invariants that make the phenomenology of two dimensional turbulence
somewhat more complex than three dimensional turbulence 
and not derivable from simple dimensional arguments.
Theoretical studies of turbulence usually employ a statistical description and often 
involve assumptions about homogeneity isotropy and the nature of the interactions.
Based on physical arguments, Kraichnan \cite{Kraichnan67}, Leith \cite{Leith68} and
Batchelor \cite{Batchelor69} conjectured that there is a dual cascade in two dimensional 
turbulence:
energy flows to the larger scales while enstrophy moves to the small scales
(when the system is driven at some intermediate scale). 
Kraichnan-Leith-Batchelor (KLB) theory assumes isotropy and homogeneity in the limit of 
infinite domain size, and in the zero viscosity limit predicts a $k^{-5/3}$ energy spectrum
for the large scales and a $k^{-3}$ energy spectrum for the small scales. 
The assumptions of the KLB theory, as well as the power index and the universality of the two
energy spectra has been questioned in the literature \cite{Kraichnan71,Saffman71,Sulem71,
Bowman96,Tran02,Gkioulekas05a,Gkioulekas05b}.

In this paper we derive some simple rigorous bounds for the long time averaged bulk energy
and enstrophy dissipation rates for two dimensional statistically stationary flows
sustained by a variety of driving forces. The study of physically relevant rigorous 
bounds on the energy dissipation rate, i.e.,
the power consumption of turbulence, for a class of boundary-driven flows can be traced
back to the seminal work of Howard \cite{Howard63},
and in recent years an alternative approach \cite{Doering92} has renewed interest in those
kinds of problems, providing direct connections to experiments in some cases.
Bounds for the energy dissipation of steady body forced flows---more convenient for
theoretical and numerical investigations---in three dimensions
have been derived by Foias \cite{Foias97} and others
\cite{Doering02,DoeringES03,Doering05,Petrov05}.
Not unexpectedly, Foias et al. \cite{Foias02} also derived a bound for the
enstrophy dissipation rate in the statistically stationary states of
two-dimensional turbulence driven by a restricted class of forces.
Bounds for the energy and enstrophy dissipation in two dimensional flow driven by
a monochromatic forcing were derived in \cite{Constantin94} and \cite{Tran02}.
The case of temporally white noise forcing was studied by Eyink \cite{Eyink96}.
More recently Tran {\it el at} \cite{Tran05,Tran06} derived bounds of the 
enstrophy dissipation for freely decaying two dimensional turbulence
in terms of the initial ideal invariants of the flow.
(See also \cite{Eyink01} for the treatment of the same problem in terms of
the inviscid Euler Equations).
Finally, we mention that bounds on the dimension 
of attractor of the 2D Navier-Stokes have been derived in 
\cite{Constantin85, Constantin88, Doering91, Doering98, Tran04b}
and more recently by \cite{Gibbon06}.

The results for the energy and enstrophy dissipation of forced flows derived in this paper
apply to a more general type of forcing than the single scale forcing
that was considered in \cite{Constantin94,Tran02}.
We also consider forces that are smoothly varying in time, unlike the white noise forcing
investigated in \cite{Eyink96}, and we are particularly interested in the behavior of 
the long times averaged dissipation rates in the vanishing viscosity limit. 
Because the viscosity is a dimensional quantity we must specify what we mean by 
``small'' viscosity. 
To be precise, we measure the magnitude of the viscosity in terms of the Reynolds
number in the statistically steady state,
\beq
Re = \frac{U}{k_{f}\nu}
\label{Re}
\eeq
where $U$ is the root mean square velocity and $k_{f}$ is a natural wavenumber 
(inverse length scale) in the driving force.
The dissipation rates are also dimensional quantities, so we measure them in terms of the 
inviscid scales determined by $U$ and $k_{f}$.
That is, we estimate
\beq
\beta = \frac{\epsilon }{k_{f}U^{3}}
\quad \quad \text{and} \quad \quad 
\gamma = \frac{\chi}{k_{f}^{3}U^{3}}
\eeq
in terms of $Re$ and focus on the $Re \rightarrow \infty$ limit holding other parameters
(such as the large scale eddy turnover time in the most general case) fixed.
For a broad class of external driving we find that
\beq
\beta \lesssim Re^{-1/2}
\quad \quad \text{and} \quad \quad 
\gamma \lesssim Re^{0},
\label{gsum}
\eeq
consistent with an enstrophy cascade of sorts.

However, for special cases of forcing such as ``ultra narrow band'' monochromatic 
(i.e., involving on a single length scale, albeit with a broad range of time dependence) 
forcing, or for a fixed energy flux forcing popular for
direct numerical simulations, a stronger bound holds:
\beq
\beta \lesssim Re^{-1}
\quad \quad \text{and} \quad \quad 
\gamma \lesssim Re^{-1}.
\label{msum}
\eeq
This kind of $Re^{-1}$ scaling suggests ``laminar'' flows where the energy is 
concentrated at or above the smallest length scale of the forcing. 
This sort of scaling has been shown before in the literature \cite{Constantin94,Tran02}
for monochromatic forcing and for white noise in time forcing \cite{Eyink96}.

In every case the bounds derived here are strictly less than those 
available --or expected-- for three dimensional turbulence.
The upper bounds (\ref{gsum}) on the energy and enstrophy dissipation for two dimensional
flows derived here are in a sense a consequence of combining the approaches in
\cite{Foias02} and \cite{Eyink96,Tran02} applied to a class of forcing 
functions concentrated in a finite range of length scales.
Even though some steps in our analysis have been taken before, in order to make this paper
self-contained the complete (but nevertheless short) proofs will be presented here.

The rest of this paper is organized as follows.  
In the next section II we introduce the problem and basic definitions, and perform the analysis
leading to (\ref{gsum}) for the simplest case of time independent body forces.
Section III generalizes the analysis to include a broad class of time-dependent forces.
In section IV we briefly review the results for 
time-dependent but single-length scale forcing 
and ``fixed-flux" forces in order to establish the stronger results in (\ref{msum}).
The concluding section V contains a brief discussion of the results and their implications.


\section{Time-independent forcing}

Consider a two dimensional periodic domain $[0,L]^2$, i.e., $\mathbb{T}^2_L$, filled with an
incompressible fluid of unit density evolving according to the Navier-Stokes equation:
\beq
\partial_t {\bf u} +{\bf u \cdot \nabla u} =-\nabla p +\nu \nabla^2{\bf u} +{\bf f}
\label{NS}
\eeq
where ${\bf u} = {\bf \hat{i}} u_x(x,y,t) + {\bf \hat{j}} u_y(x,y,t)$  is the 
incompressible (divergence-free) velocity field,
$p(x,y,t)$ is the pressure, $\nu$ is the viscosity, and
${\bf f(x,y)}= {\bf \hat{i}} f_x(x,y) + {\bf \hat{j}} f_y(x,y)$ is a smooth, mean zero,
divergence-free body force with characteristic length scale $\sim k_{f}$
(defined precisely below).
The scalar vorticity $\omega = \partial_x u_y -\partial_y u_x$ satisfies
\beq
\partial_t \omega +{\bf u \cdot \nabla}\omega = \nu \nabla^2 \omega +\phi
\label{VE}
\eeq
where $\phi = {\bf \hat{k} \cdot}\nabla \times {\bf f} = \partial_x f_y -\partial_y f_x$.

The Reynolds number is defined in (\ref{Re}) where
\beq
U \equiv \langle |{\bf u}|^{2} \rangle^{1/2}
\eeq
is the root-mean-square velocity with $\langle \cdot \rangle$ 
representing the space-time average
\beq
\langle {\bf g} \rangle = \lim_{T\to\infty} \frac{1}{T}\int_0^T 
\left(\frac{1}{L^2}\int_{\mathbb{T}^2_L} {\bf g}(x,y,t)  \, d^{2}x \right) \, dt.
\eeq 
(The limit in the time average is assumed to exist for all the quantities of interest.)
The forcing length scale associated with the wavenumber $k_{f}$ is defined by
\beq
k_f^{2} \equiv  \frac{\|\nabla^2 {\bf f}\|} {\|{\bf f}\|}
\eeq
where $\| \cdot \|$ is the $L_2$ norm on $\mathbb{T}^2_L$.
It is apparent that we are restricting ourselves to sufficiently smooth forcing functions.

The time and space averaged energy dissipation rate is
\beq
\epsilon \equiv \nu \left \langle  | {\bf \nabla u}|^2 \right \rangle
= \nu \left \langle  \omega ^2 \right \rangle,
\label{energy_diss}
\eeq
the second expression resulting from integrations by parts utilizing incompressibility.
The bulk averaged enstrophy dissipation rate is
\beq
\chi \equiv \nu \langle  |\nabla \omega| ^2 \rangle
= \nu  \langle  | {\bf \nabla^{2} u}|^2  \rangle.
\label{enstrophy_diss}
\eeq

We think of $\beta = \epsilon/k_{f}U^3$ and $\gamma = \chi/k_{f}^{3}U^3$ as functions
of $Re$ and the functional form or ``shape'' of the forcing, but not explicitly on its amplitude 
\beq
F = \frac{\|{\bf f}\|}{L}
\eeq
except indirectly through its influence on $U$.

We are considering the Reynolds number to be the ``control parameter'' even though
it is defined in terms of the emergent quantity $U$.
Strictly speaking the flow is determined by the structure and amplitude of the body force
(and the initial data) so the Grashof number such as $Gr = F/k_{f}^{3}\nu^2$ 
should naturally be used as the relevant dimensionless control parameter
indicating the intensity of the driving and the resulting flow.
Indeed, while we can always realize any given value of $Gr$, it is not at all evident that every
particular value of $Re$ can be achieved.
Nevertheless, in order to express the results in terms of quantities 
familiar to the theory of homogeneous isotropic
turbulence we will (without loss of rigor) express the bounds in terms of $Re$. 

Poincare's inequality applied to (\ref{energy_diss}) and (\ref{enstrophy_diss}) 
immediately yield the lower estimates
\beq
\epsilon \ge \nu \frac{4\pi^{2}}{L^{2}} U^{2}
\quad \quad \text{and} \quad \quad
\chi \ge \nu \frac{16\pi^{4}}{L^{4}} U^{2}
\eeq
so that
\beq
\beta \ge 4 \pi^{2} \alpha^{2} Re^{-1}
\quad \quad \text{and} \quad \quad
\gamma \ge 16\pi^{4} \alpha^{4} Re^{-1}
\eeq
where $\alpha=(k_{f}L)^{-1} \le (2\pi)^{-1}$ is the ratio of the forcing to domain length scales.
If $\beta$ and $\gamma$ scale both as $\sim Re^{-1}$ 
then we say that the flow exhibits laminar behavior
because the energy is then necessarily concentrated at 
relatively long length scales determined by the
prefactor, rather than over a broad range of scales that increases as $Re \rightarrow \infty$.

On the other hand if $\beta \sim Re^{0}$, the scaling expected in three dimensional turbulence,
the flow exhibits finite (residual) dissipation 
in the limit of zero viscosity indicating the
presence of an active and effective energy cascade to small scales.
It was recently shown \cite{Doering02, Doering05} that $\beta \le c Re^{0}$ for the
vanishing viscosity limit of three dimensional versions of the systems under consideration here
and in section III, where the coefficient $c$ is uniform in $\nu$, $L$, and $F$.
There is, however, no known {\it a priori} enstrophy dissipation rate bound for the three dimensional 
turbulence; this is related to the outstanding open question of the regularity of 
solution for the three dimensional Navier-Stokes equations.
As the results of this paper suggest quantitatively, the dissipation rates of
two dimensional turbulence falls somewhere between laminar 
scalings and the rates for three dimensional turbulence.

To prove the two dimensional bounds we first take the inner product of the vorticity equation
(\ref{VE}) with $\omega$ and average to obtain the enstrophy production-dissipation balance
\beq
\chi =  \left\langle \omega \phi \right\rangle
\label{EPDB}
\eeq
where the time derivative term drops out when we take the long time average.
Integrating by parts to move the $\hat{\bf k}\cdot\nabla\times$ from $\omega$ 
onto $\phi$ and the Cauchy-Schwarz inequality, we easily obtain
\beq
\chi \le k_{f}^{2}UF.
\label{VBI}
\eeq

For the second step, consider a smooth incompressible vector field ${\bf v}(x,y)$.
Take the inner product of $\bf v$ with the Navier-Stokes equation, integrate by parts
and average to obtain
\begin{eqnarray}
\frac{1}{L^{2}} \int_{\mathbb{T}^2_L} {\bf v\cdot f }  d^{2}x  =
- \left\langle {\bf u} \cdot (\nabla{\bf v}) \cdot{\bf u} +
\nu {\bf u} \cdot \nabla^2{\bf v}  \right\rangle.
\end{eqnarray}
Using the Cauchy-Schwarz and H\"older's inequality (as in \cite{Doering02})  we deduce
\begin{equation}
F  \times \frac{1}{L^2\|{\bf f}\|} \int_{\mathbb{T}^2_L} {\bf v\cdot f }  \, d^{2}x  \le
U^2 \| \nabla {\bf v}\|_\infty + 
\nu  \frac{U}{L} \|\nabla^2{\bf v}\|
\label{F1a}
\end{equation}
where $\|\cdot\|_\infty$ is the $L_\infty$ norm on $\mathbb{T}^2_L$.
In order for the inequality to be non-trivial we need to restrict $\bf v$ so that  
$\int_{\mathbb{T}^2_L} {\bf v\cdot f} \,  d^{2}x >0$.
This is easy to arrange.  
For example the choice  ${\bf v = f}/F$  will satisfy this condition
if ${\bf f}$ is sufficiently smooth 
that the right hand side of (\ref{F1a}) is finite.
If it is not so smooth, then for instance we can take 
${\bf v} \sim K \circ {\bf f}$ where $K(x,y,x',y')$
is a (positive) smoothing kernel. 
In any case we can choose ${\bf v}$ appropriately and use (\ref{F1a}) 
to eliminate $F$ in (\ref{VBI})
so that:
\beq
\chi \le U^3k_{f}^{3} \, \left(C_{1} +\frac{C_{2}}{Re}\right)
\quad \quad \Rightarrow \quad \quad \gamma \le \left(C_{1} +\frac{C_{2}}{Re}\right)
\label{Xbound}
\eeq
where the dimensionless coefficients $C_{1}$ and $C_{2}$
are independent of $k_f$ and $L$, depending only on the functional ``shape" of $\bf v$
(and thus also on the shape of $\bf f$) but not on its amplitude F or the viscosity $\nu$.
Explicitly they are
\begin{equation}
C_{1} = \frac {\| \nabla_l {\bf v}\|_\infty}{\langle {\bf v\cdot f}/F \rangle}
\quad \quad \mathrm{and} \quad \quad
C_{2} = \frac{\langle |\nabla_{l}^2{\bf v} |\rangle^{1/2}}{\langle {\bf v\cdot f}/F \rangle}
\end{equation}
where $\nabla_{l}$ is the gradient with respect to the non-dimensional coordinate $k_{f}{\bf x}$.
An upper bound for the enstrophy dissipation rate like that in (\ref{Xbound})
was first derived in \cite{Foias02}.
Note that for strictly band-limited forces, i.e., if the Fourier transform of the force is supported on 
wavenumbers with $|{\bf k}| \in (k_{min}, k_{max})$ with $0 < k_{min} < k_{max} < \infty$,
then the coefficients $C_{1}$ and $C_{2}$ are bounded by 
pure numbers depending only on $k_{max}/k_{min}$.

For the final step of the proof we use integrations by parts and the Cauchy-Schwarz inequality
to see that
\beq
\langle{\omega^{2}}\rangle^{2} =
\left\langle {\bf u \cdot \nabla} \times (\hat{{\bf k}} \omega) \right\rangle^{2}
\le \langle |{\bf u}|^2\rangle \langle |{\bf\nabla \omega}|^2 \rangle .
\label{trickI}
\eeq
Combining (\ref{trickI}) with (\ref{Xbound}) we deduce
\beq
\langle{\omega^{2}}\rangle^{2} \le  
\frac{k_{f} U^5}{\nu} \left(C_{1} +\frac{C_{2}}{Re} \right),
\eeq
and in terms of the energy dissipation rate this is the announced result
\beq
\beta \le Re^{-1/2} \left(C_{1} +\frac{C_{2}}{Re} \right)^{1/2}.
\label{bound}
\eeq

\section{Time-dependent forces}

Now consider the Navier-Stokes equation (\ref{NS}) where the time dependent body force
${\bf f}(x,y,t)$ is smooth and incompressible with characteristic
length scale $\sim k_{f}^{-1}$ given by
\beq
k_f^{4} \equiv  \frac{\langle |\nabla^2 {\bf f}|^2\rangle}{\langle |{\bf f}|^2 \rangle},
\eeq
and time scale $\Omega_f^{-1}$ defined by
\beq
\Omega_f^{2} \equiv \frac{\langle |\partial_t {\bf f}|^2 \rangle}
{\langle |{\bf f}|^2 \rangle}.
\eeq
We define 
\beq
\tau =  \frac{\Omega_f}{k_{f}U},
\eeq
the ratio of the ``eddy turnover'' time $(k_{f}U)^{-1}$ to the forcing time scale $\Omega_f^{-1}$.
In this time-dependent setting the amplitude $F$ of the force is
\beq
F =  \langle |{\bf f}|^{2}\rangle^{1/2}.
\eeq

As before, the space and time average of  $\omega$ times the vorticity equation (\ref{VE})
yields the enstrophy balance equation (\ref{EPDB}), and integration by parts
and the Cauchy-Schwarz inequality implies
\beq
\chi\le k_{f}^{2}UF.
\label{VB}
\eeq

For the second step here we introduce a smooth incompressible vector field 
${\bf v}(x,y,t)$ and take space and time average of the inner product of
with the Navier-Stokes equation to obtain
\begin{eqnarray}
\left\langle {\bf v \cdot f}  \right\rangle =
-  \left\langle {\bf u} \cdot \partial_{t} {\bf v}  +
 {\bf u \cdot (\nabla v) \cdot u} + \nu {\bf u} \cdot \nabla^2{\bf v} \right\rangle.
\end{eqnarray}
Cauchy-Schwarz and H\"older's inequalities then imply
\begin{eqnarray}
F \frac{ \langle{\bf v \cdot f} \rangle}{\langle |{\bf f}|^{2} \rangle^{1/2}} &\le&
U\langle |{\bf \partial_t v}|^2 \rangle^{1/2} \ + \
U^2 \sup_t \| \nabla {\bf v}\|_\infty  \nonumber \\ 
&+&
 \nu  U \langle |\nabla^2{\bf v}|^2 \rangle^{1/2}.
 \label{F1}
\end{eqnarray}
Now we need to be able to choose $\bf v$ satisfying $\langle{\bf v \cdot f} \rangle > 0$
such that all the coefficients on the right hand side are all finite.
Our ability to do this depends on details of ${\bf f}(x,y,t)$.  

For example if ${\bf f}$ is sufficiently smooth in space and appropriately uniformly
bounded in time then we can choose ${\bf v} \sim {\bf f}$.
We could also choose ${\bf v}$ as an appropriately filtered version of
${\bf f}$ to cover  more general cases.
For the purposes of this study and to display the results in the clearest 
(if not the sharpest or most general form) 
we will simply presume that ${\bf f}$ is sufficiently regular
that we can take  ${\bf v} = {\bf f}$.
In that case (\ref{F1}) becomes
\beq
F \le \Omega_{f}U +
U^2 \, \frac{\sup_t  \| \nabla {\bf f}\|_\infty}{F}  +
\nu k_{f}^2U.
 \label{F2}
\eeq
Then using this to eliminate $F$ from (\ref{VB}) we have
\beq
\chi \le k_{f}^{3}U^3 \left( \tau + C_{3} +\frac{1}{Re} \right)
\ \  \Rightarrow \ \ \gamma \le 
\left( \tau + C_{3} +\frac{1}{Re} \right)
\label{VboundT}
\eeq
where the coefficient $C_{3}$ is
\beq
C_{3} = \frac{\sup_t \| \nabla_{l} {\bf f}\|_\infty}{F}
\eeq
with $\nabla_{l}$ denoting the gradient with respect to the non-dimensional coordinate 
$k_{f}{\bf x}$.
The dimensionless number $C_{3}$ is independent of the scales of $F$, $k_f$, $L$, etc.,
depending only on the ``shape" of $\bf f$.
For example if  $\bf f$ is quasi-periodic with $N$ frequencies and involves only wavenumbers
${\bf k}$ with $0 < k_{min} < |{\bf k}| < k_{max} < \infty$,
then $C_{3}$ is bounded by $\sqrt{N}$ times a function of $k_{max}/k_{min}$.

The final step again uses the inequality
\beq
\epsilon^{2} = \nu^{2} \langle{\omega^2}\rangle^{2} =
\nu^{2}\langle {\bf u} \cdot \nabla \times (\hat{{\bf k}} \omega) \rangle^{2} 
\le \nu U^{2} \chi
\label{trick2}
\eeq
and it then follows immediately from (\ref{VboundT}) that
\beq
\beta \le Re^{-1/2} \left(\tau + C_{3} +\frac{1}{Re} \right)^{1/2}.
\label{boundT}
\eeq
Note in this case $\tau$ depends on $U$ and features of the forcing through
$k_{f}$ and $\Omega_{f}$, but {\it not} on $\nu$.

\section{Monochromatic and constant flux forces}

An even sharper scaling bound on the energy and enstrophy dissipation rates can be derived when
the driving is monochromatic in space, whether it is steady or time dependent
\cite{Constantin94,Tran02}. 
Suppose the body force involves only a single length scale, i.e.,
\beq
-\nabla^{2} {\bf f} =k_{f}^{2}  {\bf f}.
\eeq
This does not preclude complex time-dependence for ${\bf f}(x,y,t)$, just that it
involves only spatial modes with wavenumbers ${\bf k}$ with $|{\bf k}| = k_{f}$.
Then the enstrophy production-dissipation balance (\ref{EPDB}) implies
\beq
\chi =  \langle \omega \phi \rangle =
\langle {\bf u} \cdot (-\nabla^{2}{\bf f}) \rangle = 
k_{f}^{2}\langle {\bf u} \cdot {\bf f} \rangle = k_{f}^{2} \epsilon.
\label{EPDBmono}
\eeq
Combining this with (\ref{trick2}), we observe that
\beq
\epsilon^{2} \le \nu U^{2} \chi = \nu k_{f}^{2}U^{2}\epsilon
\eeq
so that
\beq
\epsilon \le \nu k_{f}^{2} U^{2}
\quad \quad \text{and} \quad \quad 
\chi \le \nu k_{f}^{4} U^{2}
\label{trick3}
\eeq
implying that both $\beta$ and $\gamma$ are bounded by $Re^{-1}$.
Note that this kind of monochromatic forcing is a special case that
has been shown in the literature for some cases 
to lead to a laminar flow that never looses stability
\cite{Marchioro86}.

An other type of forcing that results in this scaling is
\beq
{\bf f}(x,y,t) = \epsilon \frac{{\cal P}{\bf u}}{L^{-2}\|{\cal P}{\bf u} \|^{2}} 
\eeq
where ${\cal P}$ is the projector onto spatial modes of wavenumber
${\bf k}$ with $|{\bf k}| \in [k_{min}, k_{max}]$,
and the coefficient $\epsilon$ is now the control parameter.
This type of forcing is often applied in
numerical simulations of homogeneous isotropic turbulence.
With this forcing in the Navier-Stokes equations constitutes an autonomous dynamical system
with kinetic energy injected at a constant rate $\epsilon$ at wavenumbers with 
$|{\bf k}| \in [k_{min}, k_{max}]$.
The rms speed $U$ (i.e., the Reynolds number) and the enstrophy dissipation rate $\chi$
are then emergent quantities, determined by the imposed energy flux $\epsilon$.
The mean power balance for solutions is still
\beq
\nu \langle |\nabla {\bf u}|^{2} \rangle = 
\nu \langle \omega^{2} \rangle = \epsilon,
\eeq
and the enstrophy production-dissipation balance reads
\beq
\chi =  \nu \langle |\nabla {\omega}|^2 \rangle
= \epsilon 
\left\langle \frac{\|\nabla {\cal P}{\bf u}\|^{2}}{\|{\cal P}{\bf u} \|^{2}} 
\right\rangle.
\label{EPDBff}
\eeq
Because the the forcing only involves wavenumbers in $[k_{min}, k_{max}]$
with positive energy injection at each wavenumber, at each instant of time
\beq
 k_{min}^{2} \|{\cal P}{\bf u} \|^{2} \le \|\nabla {\cal P}{\bf u}\|^{2}
 \le k_{max}^{2} \|{\cal P}{\bf u} \|^{2}.
\eeq
Then (\ref{EPDBff}) implies that
\beq
k_{min}^{2} \epsilon \le \chi \le k_{max}^{2} \epsilon.
\label{con}
\eeq
Using this with (\ref{trick2}) we see that 
\beq
\epsilon^{2} \le \nu U^{2} \chi \le \nu U^{2}  k_{max}^{2} \epsilon,
\eeq
and we conclude
\beq
\epsilon \le \nu k_{max}^{2} U^{2}
\quad \quad \text{and} \quad \quad 
\chi \le \nu k_{max}^{4}U^{2}.
\eeq
Hence also in this case both $\beta$ and $\gamma$ are bounded $\sim Re^{-1}$.

Note that in both these derivations a condition like (\ref{con}) or the
stronger condition (\ref{EPDBmono}) was used. 
It is an open question whether such a condition holds for more general 
and more ``realistic'' forcing functions.
The results (\ref{VboundT}) and (\ref{boundT}) give restrictions 
on the energy and enstrophy dissipation rate for a broader
class of driving, but it is natural to wonder how
broad of a class of forcing functions
would actually result in the
$Re^{-1}$ scalings
in the vanishing
viscosity
limit.

\section{Discussion}

These quantitative bounds show that for two dimensional 
turbulence sustained by forces as described
in the previous sections, there is no residual dissipation in the 
vanishing viscosity limit defined
by $Re \rightarrow \infty$ at fixed $U$, $L$, $k_{f}$ and $\Omega_{f}$.
To be precise, $\epsilon$ vanishes at least as fast as $Re^{-1/2}$ in this limit.
This confirms that there is no forward energy cascade
in the steady state in the inviscid limit.
On the other hand the residual enstrophy dissipation allowed by (\ref{VboundT}) in this limit
does not rule out a forward enstrophy cascade.
This combination, $\epsilon \rightarrow 0$ with $\chi = {\cal O}(1)$ in the inviscid limit, is
consistent with the dual-cascade picture of two-dimensional turbulence developed by
Kraichnan \cite{Kraichnan67}, Leith \cite{Leith68} and Batchelor \cite{Batchelor69}.
Note however that the absence of forward energy cascade is not necessarily true  for any
finite value of $Re$. 
The $\beta\sim Re^{-1/2}$ scaling allowed by the bound is less severe than what a laminar flow 
(or a flow with only inverse cascade of energy) would predict, and as a result 
(\ref{bound}) does not exclude the presence of some direct subdominant cascade
of energy when the Reynolds number is finite as suggested by \cite{Gkioulekas05a}.

On the other hand the direct cascade of enstrophy is necessarily absent 
for some forcing functions (see \cite{Constantin94,Eyink96,Tran02}).
When the forcing acts at a single scale or constant power is injected in a 
finite band of wavenumbers, both $\epsilon$ and $\chi$ vanish $\sim Re^{-1}$.
This suggests an essentially laminar behavior for these flows: if the energy spectrum 
follows a power law $E(k)\sim k^{-\alpha}$ for large wavenumbers then the exponent
must be $\alpha \le -5$ for $\chi$ to vanish in the vanishing viscosity limit.
These results have been interpreted as absence of enstrophy cascade in finite domains. 
However, both of these results rely on the condition (\ref{con})
which is not guaranteed for a general forcing functions.
Whether (\ref{con}) might hold for more general forcing functions is an open question;
the results (\ref{VboundT}) and (\ref{boundT}) give the restrictions on the
energy and enstrophy dissipation rate for a general forcing.
Note that the $\beta\sim Re^{-1/2}$ does not impose significant restriction on the
energy spectrum given the bound on $\chi$.
These considerations suggest that it is possible therefore that in two dimensional 
turbulence the steady state energy spectrum depends on the type of forcing used,
even within the class of relatively narrow-band driving.
High resolution numerical simulations with forcing that
does not necessarily satisfy (\ref{con})
would be useful at this point to resolve this issue.

We conclude by noting that an interesting question that follows from these results is that of
the $Re$-scaling of the energy dissipation in systems that almost have two dimensional 
behavior like strongly rotating, strongly stratified or conducting
fluids in the presence of a strong magnetic field. 
For example, is there a critical value of the rotation 
such that the scaling of the energy dissipation 
rate with the Reynolds number transitions from $\epsilon \sim Re^0$ 
to $\epsilon \sim Re^{-1/2}$?
These questions remain for future studies.   

\begin{acknowledgements}

The authors thank J.D. Gibbon and M.S. Jolly for helpful discussions and comments.
CRD was supported in part by NSF grants PHY-0244859 and PHY-0555324
and an Alexander von Humboldt Research Award.
AA acknowledges support from the National Center for Atmospheric Research.
NCAR is supported by the National Science Foundation. 
\end{acknowledgements}

\bibliographystyle{unsrt}
\bibliography{ms}

\end{document}